# Non-Fungible Token Security

A deep dive into this new form of cryptocurrency and art trading.

Ryleigh McKinney, Sundar Krishnan
Angelo State University
rmckinney5@angelo.edu, skrishnan@angelo.edu

*Abstract*— **Non-fungible tokens (NFTs) are unique digital assets stored on the blockchain and is used to certify ownership and authenticity of the digital asset. NFTs were first created in 2014 while their popularity peaked between 2021 and 2022. In this paper, the authors dive into the world of Non-Fungible Tokens (NFTs), their history, the Future of NFTs, as well as the security concerns.**

## INTRODUCTION

What is a Non-Fungible Token? Also known as an NFT, Non-fungible tokens are cryptographic assets on a blockchain with unique identification codes and metadata that distinguish them from each other.[1] In more simple terms, an NFT is a graphic design that may have an animation, they can be traded and bought for real money. NFTs are exclusive and the designs are only made once. They can also represent real things such as artwork and even real estate. Some people use NFTs as a form of self-expression.

## HISTORY OF NFTs

The "first" NFT was created in 2012, so, NFTs have been around a lot longer than people have realized. From 2012-2016, NFTs simply began as colored coins. The idea of Colored Coins was to describe a class of methods for representing and managing real-world assets on the blockchain to prove ownership of those assets; like regular Bitcoins, but with an added 'token' element that determines their use, making them segregated and unique.[2] These colored coins laid down the foundation for NFTs.

On May 3rd, 2014, digital artist Kevin McCoy minted the first-known NFT 'Quantum' on the Namecoin blockchain. 'Quantum' is a digital image of a pixelated octagon that hypnotically changes colour and pulsates in a manner reminiscent of an octopus.[2] Around the years of 2017-2020, that is when NFTs truly started to take off into the investment and stock world, otherwise known as going mainstream. The big shift for NFTs to Ethereum was backed up with the introduction of a set of token standards, allowing the creation of tokens by developers. The token standard is a subsidiary of the smart contract standard, included to inform developers how to create, issue and deploy new tokens in line with the underlying blockchain technology.[2] After this rise, Vancouver-based venture studio Axiom Zen introduced CryptoKitties. CryptoKitties is a virtual game based on the Ethereum blockchain. The game enables players to adopt, breed and trade virtual cats, storing them in crypto wallets. After its announcement it wasn't long before the game became a viral sensation, becoming so popular that CryptoKitties clogged the Ethereum blockchain and people began making unbelievable profits.[2] After the huge success of CryptoKitties, NFT gaming grew to be more popular as the years went by. 2021 was the year NFTs began booming, causing more supply and demand for these cryptographics.

One of the biggest factors in this boom was the huge changes that occurred within the art market and the industry at large, when prestigious auction houses; Christie's and Sotheby's namely, not only took their auctions into the online world but also began selling NFT art.[2] Christie was able to sell their NFT, Beeple's

Everydays: the First 5000 Days, for $69 million. Soon after this sale, other platforms began creating and marketing their own versions of NFTs. This included blockchains such as Cardano, Solano, Tezos and Flow. With these newer platforms for NFTs, some new standards were established to ensure the authenticity and uniqueness of the digital assets created.[2] Near the end of 2021, Facebook rebranded as Meta and moved into the metaverse. This changed into a new universe surged the demand for NFTs even more.

Figure 1 shows "Everydays" as the First 5000 Days is a digital work of art created by Mike Winkelmann, known professionally as Beeple. The work is a collage of 5000 digital images created by Winkelmann for his Everydays series. Its associated non-fungible token (NFT) was sold for $69.3 million at Christie's in 2021, making it first on the List of most expensive non-fungible tokens.[3]

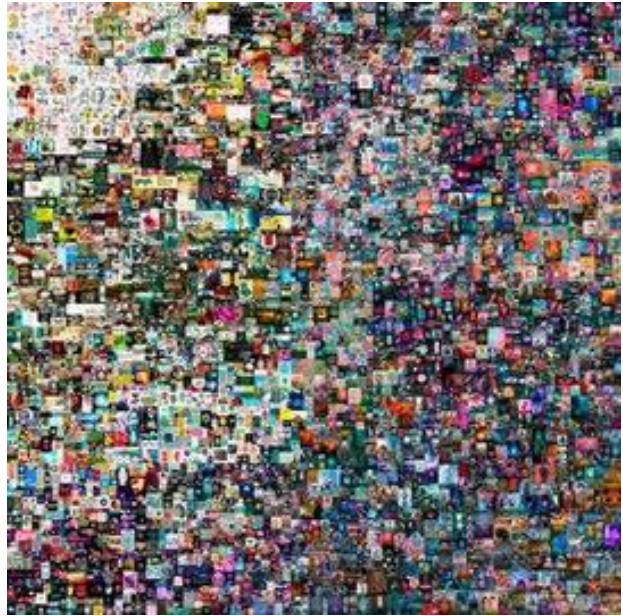

Figure 1: The first 5000 Days[7]

THE FUTURE OF NFTS

The introduction to NFTs was simply just the beginning, there is much more that can be done to expand and grow this new form of cryptocurrency. NFTs will have a big impact on the world of Gaming, Social Impacts, Real Estate, and The Metaverse. Are NFTs going to have more of an impact than people realize? The creation of CrytoKitties was simply the beginning, people tend to be more attracted to games were collecting items and building collections is the main focus point. They feel a sense of need and want to play, to continue to build their collection and be better than their friend or family member. Gamers find deep intrinsic value in their digital identities; their personal history, achievements, communities, stories, and status.[4] All Home Connections surveyed 1,000 American gamers to find out how much money they spend every month and how old they are.[5] The results are shown in Figure 2.

In one lifetime, it is shown that Gen Z and Millennials will spend as much as the average American's salary in one lifetime. This study has further proved the need for identity and individuality in the gaming community. Creators have expanded the NFT community by creating their own applications and websites to encourage identity, as well as even to encourage public good within the NFT Community. An example of this would be Jeremy Dela Rosa, the founder of Leyline, a non-profit organization that is leading the charge with social impact through NFTs. Leyline's mission is to create a sustainable NFT identity and ecosystem that celebrates, rewards, and gamifies social and environmental good.[4]  Users on their platform earn NFT collectables by doing positive deeds in the world. Their NFTs represent a cause or goal that people worked hard for, a unique moment in time, that made the world a better place.[4] Leyline creates a source of business, which was the original idea of blochain companies creating NFTs.

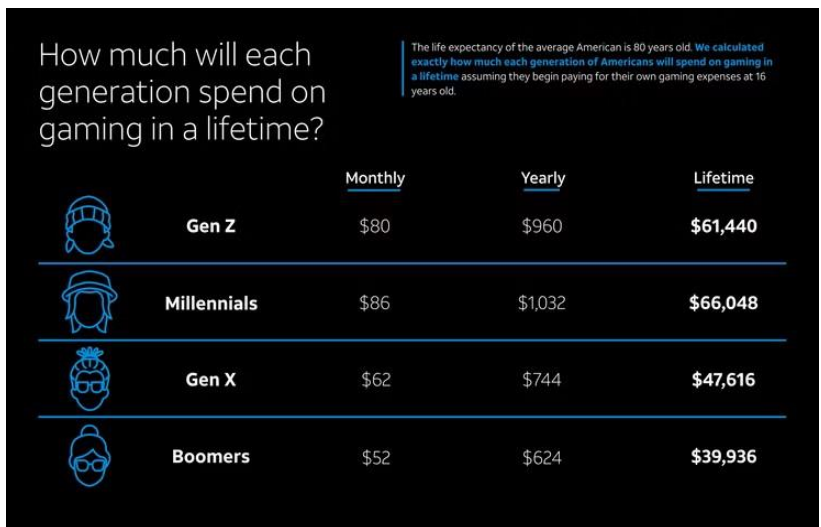

Figure 2: Cost of gaming[5]

Real estate markets are ripe for disruption. Contracts, certifications, ownership and claim history will all be stored in the blockchain and publicly accessible.[4] Selling NFTs as a form of real estate can eliminate some of the duller points in purchasing real estate, such as piles of long, and boring paperwork. As well as it can mitigate Fraud. Once more real estate contracts get created as NFTs and stored on a secure blockchain, real estate fraud will be a thing of the past, as such Smart Contracts are nearly impossible to alter, easy to verify and permanently stored. Any digital asset created as an NFT enjoys this level of security.[4] Futurist and disrupter, gmoney adds, "Mortgages are also NFT's. Would the 2008 crisis have even happened if all the MBS indexes were fully transparent, on chain? There would have not been any possibility for re-hypothecation, underlying assets and leverage would have been able to be monitored in real-time, and the entire financial system wouldn't have come crumbling down, needing a bailout by taxpayers."[4]

Finally, the Metaverse. A large and new world of virtual reality. Metaverses are virtual worlds where essentially the internet is brought to life. There is an abundant possibility within metaverse worlds, where you get to design your life, interact with real people in virtual communities, design your avatar, work, play, and explore new worlds by using virtual reality headsets, augmented reality glasses, smartphone apps or other devices which include incorporating virtual touch.[4] Entire virtual worlds are being created and built every single day. The rise in popularity of virtual reality is inevitable, especially if virtual reality is very similar to the real world. With the creation of NFT virtual worlds, buyers and creators will be able to interact with their purchases and creations, making them feel more real. Attracting people to the idea of NFTs they can interact with, and maybe even speak to through voice chat. One example of this is what is known as a "Sandbox" NFT game, it is an open world where you can do and create anything the users heart desires.

### SECURITY CONCERNS OF NFTS

With all things that seem great and perfect, there are always consequences that come with it. NFTs are not the exception, especially in a world where the cyber world is a dangerous place and can cause great damage to someone's device, or even their life. Go Banking Rates[8] identifies five main security concerns for NFTs, and how they can be dangerous.

The first one that is listed is Traditional Phishing Scams. Phishing is the fraudulent practice of sending emails purporting to be from reputable companies to induce individuals to reveal personal information, such as passwords and credit card numbers.[6] In about three hours in February 2022, more than $1.7 million in NFTs were stolen from OpenSea users via a simple phishing attack. Users were asked to sign online contracts allowing them to trade tokens, but vital portions of the authorizations were left blank. This allowed scammers to complete the forms and transfer NFT ownership from the original users.[7] As shown by the example above, scammers can disguise themselves as an NFT artist or trader to get sensitive information from users. To be more specific, a scammer could use spear phishing, which is a form of phishing used to attack a certain group or person. For example, someone could be looking for a specific NFT, a scammer could email that person advertising that they're willing to sell that specific NFT for a price. A spear phisher could get that person's card information, date of birth, and even social security if that person is not paying attention. Someone who doesn't know how to identify potential phishing scams could end up losing more money than they want to and could even have their sensitive personal information leaked.

In the world of NFTs, there is no reliable way to see the reliability of Marketplaces. Although NFT markets now trade billions of dollars worth of NFTs, both the asset class and the marketplaces themselves are just in their infancy. While security protocols are in place, hackers are constantly looking for sources of weakness.[6] With these marketplaces, they store a lot of vital information about users. If a user uses the same password on the marketplace as they do on other platforms, the likelihood of information being stolen rises. These marketplaces are very susceptible to an attack. It is important for these marketplaces to store their data safely and ensure the security of a user's personal information but, that doesn't mean these marketplaces will do that.

The next concern is the outright theft of tweets. One of the most fascinating aspects of the NFT market is that nearly anything can be turned into something of value.[6] One scam involves a tweet bot that automatically converts tweets into NFTs, which scammers could then immediately claim ownership over. In essence, if someone posts their own original work of art as a tweet, they could lose control over whatever they post if a scammer gets hold of it.[6] Theft of content is unethical, especially if that content does not have a patent or trademarks protecting the originality of the idea. In the world of Twitter and social media, content is constantly stolen. It is re-made in various forms but, the selling of this content can make it more difficult to be original and even create content.

The loss of title and ownership records is another concern. NFTs don't have traditional paper trails that you can follow to prove ownership of your asset. When you buy a home, for example, legal title passes to you and is filed as a public record.[6] So, in all reality you could own the NFT one day, and the next day you aren't the owner anymore. This can cause a loss of money spent to get that NFT, especially if it was a more expensive one. As well as fights, which could resort to real-life violence. You never truly "own" an NFT, you simply have a license for it and a right to display it.

Finally, legal issues can come into play. People can create illegal copies of the artwork. If you buy an NFT of a work of art, you don't physically take possession of the actual artwork. In fact, you likely don't even "own" the underlying art, just the digital image that you purchased.[6] This can make it hard to know if you even own the original of the artwork. For example, if you were rich enough to buy the original "Starry Night" by Vincent Van Gough, you would know it is the original that was made, yes there are copies of it but, there is truly only one original of the painting.

## CONCLUSION

In conclusion, NFTs are becoming a more popular form of cryptocurrency. They are the future of many things such as the metaverse, real estate and so much more. It is a very interesting and unique world full of more history than we realized, and that even I, as the author realized. There are various factors that play into this creation that make it so unique and interesting to study. While I personally would not invest in this, I found that it was interesting to learn more about it and see how much of an impact it has on the world. New NFTs are being created every single day as well as being purchased every single day. It is a never-ending market that will only continue to grow and franchise. There is no doubt in my mind that this form of cryptocurrency will continue to grow, it is something that may never go away due to the uniqueness of the idea. While there are some security concerns, they are security concerns you are likely to experience in the real world.